# Oxygen-Terminated (1 × 1) Reconstruction of Reduced Magnetite $Fe_3O_4$(111)

*Florian Kraushofer[1,†], Matthias Meier[1,2], Zdeněk Jakub[1,‡], Johanna Hütner[1], Jan Balajka[1], Jan Hulva[1], Michael Schmid[1], Cesare Franchini[2,3], Ulrike Diebold[1], Gareth S. Parkinson[1]\**

[1] Institute of Applied Physics, Technische Universität Wien, Wiedner Hauptstraße 8-10/E134, 1040 Wien, Austria

[2] University of Vienna, Faculty of Physics and Center for Computational Materials Science, 1090 Wien, Austria

[3] Alma Mater Studiorum, Università di Bologna, 40127 Bologna, Italy

AUTHOR INFORMATION

**Corresponding Author**

\* parkinson@iap.tuwien.ac.at

**Present Addresses**

† Department of Chemistry, School of Natural Sciences, Technical University of Munich, 85748 Garching, Germany

‡ CEITEC – Central European Institute of Technology, Brno University of Technology, 61200 Brno, Czech Republic

**ABSTRACT**. The (111) facet of magnetite ($Fe_3O_4$) has been studied extensively by experimental and theoretical methods, but controversy remains regarding the structure of its low-energy surface terminations. Using density functional theory (DFT) computations, we demonstrate three reconstructions that are more favorable than the accepted $Fe_{oct2}$ termination in reducing conditions. All three structures change the coordination of iron in the kagome $Fe_{oct1}$ layer to tetrahedral. With atomically-resolved microscopy techniques, we show that the termination that coexists with the $Fe_{tet1}$ termination consists of tetrahedral iron capped by three-fold coordinated oxygen atoms. This structure explains the inert nature of the reduced patches.

**TOC GRAPHICS**

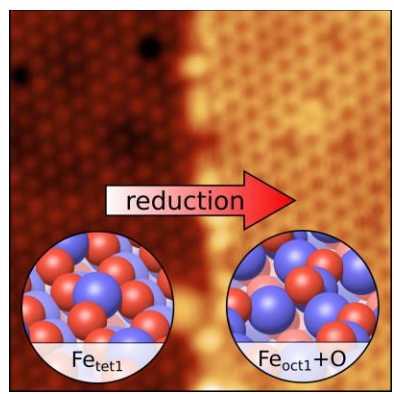

Magnetite ($Fe_3O_4$) is extremely common in nature and an important catalyst material.[1-3] While the surface structure of the (001) facet is mostly well-understood,[3, 4] the lowest-energy $Fe_3O_4$(111) surface remains controversial despite decades of study. A key issue is that multiple terminations often coexist depending on both the preparation conditions and sample history. This complicates the interpretation of area-averaging methods,[5, 6] and necessitates the use of local probes such as scanning tunneling microscopy (STM). Many atomically resolved STM images of UHV-prepared samples have been published, but questions remain, particularly about the structures formed in reducing conditions.

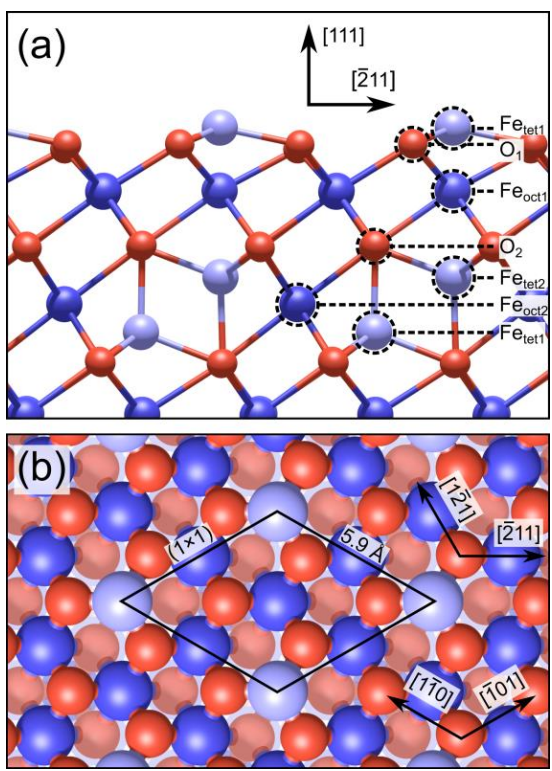

**Figure 1.** The $Fe_3O_4$(111) $Fe_{tet1}$ termination in (a) side and (b) top view. Tetrahedrally coordinated iron is light blue, octahedrally coordinated iron is dark blue, and oxygen is red. Oxygen in the deeper $O_2$ layer is pale red in (b). The layer naming convention is indicated in (a), and a (1 × 1) unit cell is drawn in (b).

Samples annealed in oxygen-rich conditions ($p_{O2} \approx 10^{-6}$ mbar, T = 870-1000 K[7-11]) usually exhibit a hexagonal array of protrusions with a nearest neighbor distance of 5.9 Å. Today, it is generally accepted that this corresponds to a relaxed bulk-truncation at the $Fe_{tet1}$ plane (see Figure 1 for layer labelling and a top view of the $Fe_{tet1}$ structure).[7-9, 12] This surface typically coexists with areas of a second (1 × 1)-periodic honeycomb structure. This has been attributed to an $Fe_{oct2}$ termination,[11, 13] which DFT calculations suggest becomes more stable than the $Fe_{tet1}$ termination under reducing conditions. A long-range ordered structure known as the "biphase" termination emerges at extremely reducing conditions, and this has been interpreted as either islands of $Fe_{1-x}O(111)$ coexisting with magnetite[14-17] or as a moiré pattern formed by an FeO-like terminating layer.[18]

In this letter, we introduce a revised phase diagram of $Fe_3O_4(111)$ featuring three new terminations that are more stable than the $Fe_{oct2}$ surface under reducing conditions. On the basis of noncontact atomic force microscopy (ncAFM) images, we assign the honeycomb patches observed experimentally to a termination at the $Fe_{oct1}$ plane, with an additional oxygen capping layer.

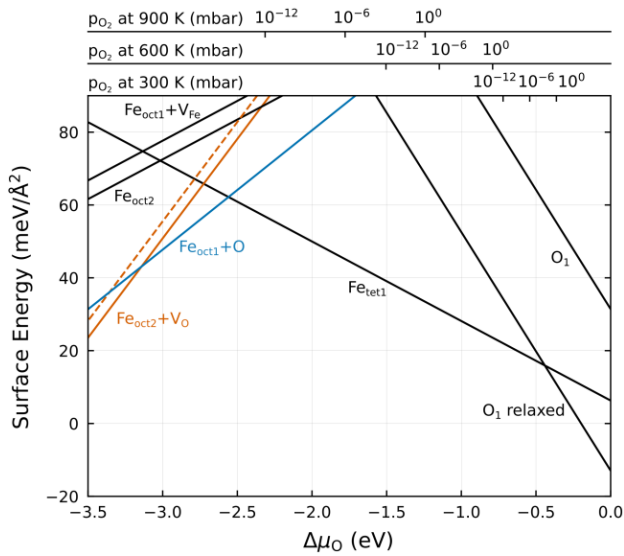

**Figure 2.** Surface energies of different terminations as a function of oxygen chemical potential $\Delta\mu_O$. The top axes indicate the corresponding oxygen partial pressures at three temperatures. Colored lines are new models introduced here, black lines correspond to terminations considered in previous work. "$O_1$ relaxed" is the modified $O_1$ termination introduced in ref. [10], while all other models can be found in ref. [19]. The dashed orange line corresponds to the $Fe_{oct2}+V_O$ termination with registry shift, as discussed in the text.

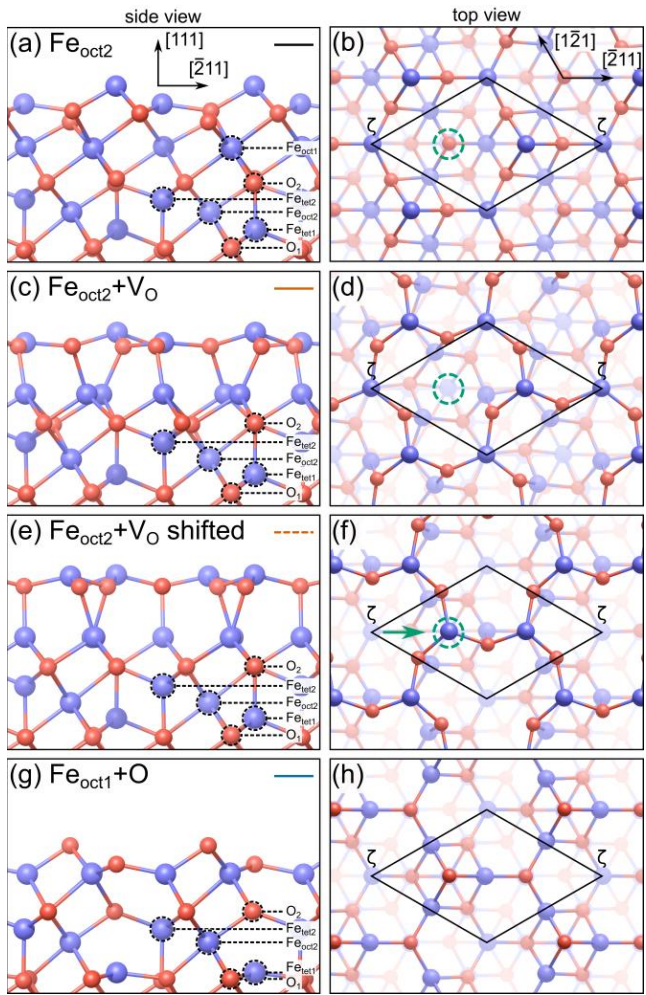

**Figure 3.** Reduced terminations of the $Fe_3O_4$(111) surface. Iron is blue (large), oxygen is red (small). (a, b) The "standard" $Fe_{oct2}$ termination. (c, d) The $Fe_{oct2}$ termination with one additional oxygen vacancy at the site marked by a dashed green circle in (b, d). (e, f) Registry-shifted version of the $Fe_{oct2}+V_O$ structure, obtained by moving one surface iron as indicated by the green arrow in (f). (g, h) Relaxed $Fe_{oct1}$ termination with iron trimers capped by an additional oxygen atom per unit cell. A (1 × 1) unit cell is indicated in black, with the corners at $Fe_{tet1}$ positions (labelled as site ζ, see below). Line styles corresponding to Figure 2 are shown in the top-right corners of (a, c, e, g).

Figure 2 shows the updated surface phase diagram of $Fe_3O_4$(111) based on our DFT+$U$ calculations. Black lines correspond to the most favorable terminations published previously,[10, 19] and colored lines correspond to the new models introduced here. We find three (1 × 1)-periodic reconstructions to be favorable over the existing models in reducing conditions. The corresponding atomic models are shown in Figure 3.

The first structure (corresponding to the solid orange line in Figure 2) is essentially an Fe$_{oct2}$ termination with one additional surface oxygen vacancy per unit cell, as shown in Figure 3 (c, d). The vacancy position is indicated by a dashed green circle in Figure 3 (b, d). This modification yields lower surface energies than Fe$_{oct2}$ in all conditions where a reduced termination is favorable over the Fe$_{tet1}$ surface. Upon creation of the vacancy, the remaining surface oxygen atoms relax outwards and each breaks one bond to an underlying Fe$_{oct1}$ atom. This leaves the subsurface iron layer tetrahedrally coordinated. The reduced coordination of surface oxygen allows rotation of the surface FeO$_3$ moieties, reducing the plane symmetry group from $p3m1$ to $p3$. As a result, the top Fe$_{oct2}$ atoms gain the necessary space to relax further into the surface, forming an almost planar Fe$_2$O$_3$ layer. We denote this structure as the "Fe$_{oct2}$+V$_O$" termination.

The surface oxygen vacancy and the reduced coordination of the surface Fe$_2$O$_3$ layer facilitate a further modification, shown in Figure 3 (e, f). The Fe$_{tet1}$ atom [positioned at the unit cell corners in Figure 3 (d)] can be moved laterally into the oxygen vacancy site, as indicated by the green arrow in Figure 3 (f). The shift enables further relaxation of the surface, and allows the subsurface iron tetrahedra to become less distorted. Nevertheless, this configuration (dashed orange line in Figure 2) is energetically less favorable than the Fe$_{oct2}$+V$_O$ without the registry shift. However, the energy difference depends somewhat on the theoretical setup (Figure S1): Using experimental Fe$_3$O$_4$ lattice constants ($a$ = 5.94 Å), the registry shift would cost ≈ 5 meV/Å$^2$, but this value is reduced to only ≈ 2 meV/Å$^2$ for a slab constructed from a PBE+$U$-optimized bulk ($a$ = 5.98 Å). The difference is likely due to an increased sensitivity of the surface Fe$_2$O$_3$ layer to strain: In all other structures considered here, sub-optimal Fe–O distances due to in-plane strain can be compensated by expanding the structure in the out-of-plane direction with only minor changes to each atom's environment. In contrast, the Fe$_{oct2}$+V$_O$ structure seems to favor coplanar iron and oxygen in the topmost layer, and reducing the lattice constant forces at least one iron atom farther out of the surface. In summary, the DFT results suggest that the registry shift is unfavorable, but the energetic differences are too small to unambiguously rule out either model. As will be discussed below, however, the Fe$_{oct2}$+V$_O$ model is in conflict with experimental data without the registry shift.

Finally, we report another competitive reduced reconstruction based on adding one oxygen atom per unit cell to the $Fe_{oct1}$ termination. The structure after relaxation is shown in Figure 3 (g, h), and a more comprehensive illustration of the relaxation and the spatial relationship to the $Fe_{tet1}$ termination is given in Figure S2. Importantly, in addition to the capping oxygen atom, a subsurface oxygen atom breaks a bond to a subsurface $Fe_{tet1}$ atom and relaxes to a 3-fold coordinated bridging site, such that the surface is terminated by two oxygen atoms per unit cell. This leaves one subsurface iron atom under-coordinated (three O neighbors), but this is compensated by the resulting near-perfect tetrahedral coordination of the three surface iron atoms. Despite being oxygen-terminated, this termination is still reduced with respect to bulk $Fe_3O_4$. All iron atoms in the surface Fe layer (formally $Fe_{oct1}$, now tetrahedrally coordinated) exhibit a Bader charge of +1.28 e. In bulk-like layers, we find a charge disproportionation of ≈0.3 e resulting in $Fe^{2+}$-like and $Fe^{3+}$-like octahedral iron with Bader charges of 1.37-1.39 e and 1.67-1.70 e, respectively, in good agreement with previous results for bulk magnetite.[20] For $Fe_{tet}$ ions in bulk-like layers, which should always be in a 3+ state, we find a Bader charge of 1.62 e. Therefore, we assign the surface $Fe_{oct1}$ cations to be in a 2+-like charge state. Interestingly, a similar $Fe_{oct1}$+O model (without relaxation) was previously proposed by Lennie et al. for what is now considered the $Fe_{tet1}$ termination,[13] but was subsequently discarded.

STM and ncAFM experiments were performed to complement the computational results. To ensure that our sample preparation yields surfaces comparable to the most recent literature, we first prepared a homogeneous (1 × 1)-periodic surface, corresponding to the previously reported $Fe_{tet1}$ termination.[7, 8, 11] Samples were sputtered (1 keV $Ar^+$ ions, 10 minutes) and annealed for 15 min in $10^{-6}$ mbar $O_2$ at 870-930 K, and then kept at the annealing temperature for another 5 min after evacuating $O_2$ to ensure low residual oxygen pressure during cooling. This avoids the formation of oxygen-related defects.[7] The best annealing temperature for producing defect-poor surfaces varied from sample to sample, most likely due to the fact that our thermocouples were not mounted directly on the samples, causing some systematic error.

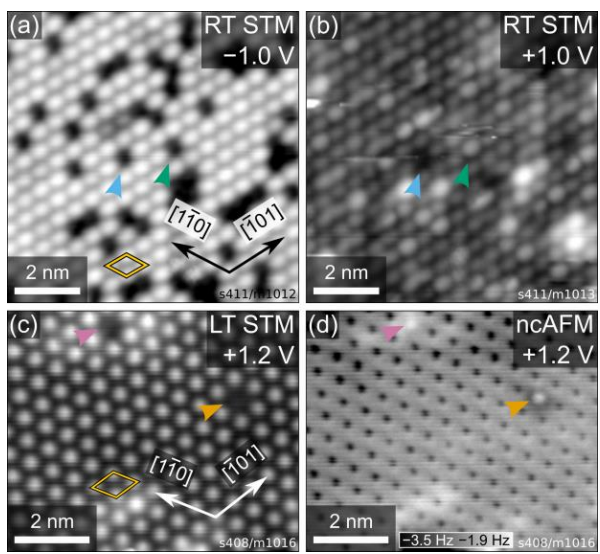

**Figure 4.** STM and ncAFM images of the $Fe_3O_4(111)$ $Fe_{tet1}$ termination. (a, b) Consecutive room-temperature STM images ($I_{tunnel} = 0.1$ nA) showing the same sample area, imaging (a) filled states ($U_{sample} = -1.0$ V) and (b) empty states ($U_{sample} = +1.0$ V). (c, d) Constant-height STM and ncAFM images acquired simultaneously at $LN_2$ temperature with $U_{sample} = +1.2$ V.

We then performed STM and ncAFM (Figure 4), as well as water temperature programmed desorption (TPD) measurements (Figure S3) to confirm that our preparation of single crystal surfaces yields the same $Fe_{tet1}$ termination as the thin film growth reported in ref. [21]. Both the STM images and the water TPD correspond well to previously published data.[11, 13, 21] Bright features in STM are attributed to $Fe_{tet1}$ atoms. The missing features have been attributed to adsorbates in cases where no feature is missing in empty-states STM, and to Fe vacancies where features are missing in both filled and empty states STM.[11] However, low-temperature STM and corresponding ncAFM images [Figure 4 (c, d)] show that apparent vacancies which are seemingly identical in STM can also differ: The defect marked with the magenta arrow appears to show weak interaction in ncAFM, which may correspond to an iron vacancy. The vacancy-like feature in STM marked by the orange arrow shows different interaction in ncAFM, possibly due to an adsorbate.

Next, we address the termination frequently found to coexist with $Fe_{tet1}$ areas on slightly reduced samples, which has previously been assigned as an $Fe_{oct2}$ termination.[11, 13] Since our DFT results indicate that this assignment is likely incorrect, we will here refer to it simply as the "honeycomb termination" when describing experimental evidence, based on its appearance in scanning probe images. Note that this is different from the "biphase" reconstruction, which also has a honeycomb appearance, though at a much larger scale (≈ 5 nm periodicity).[14-17]

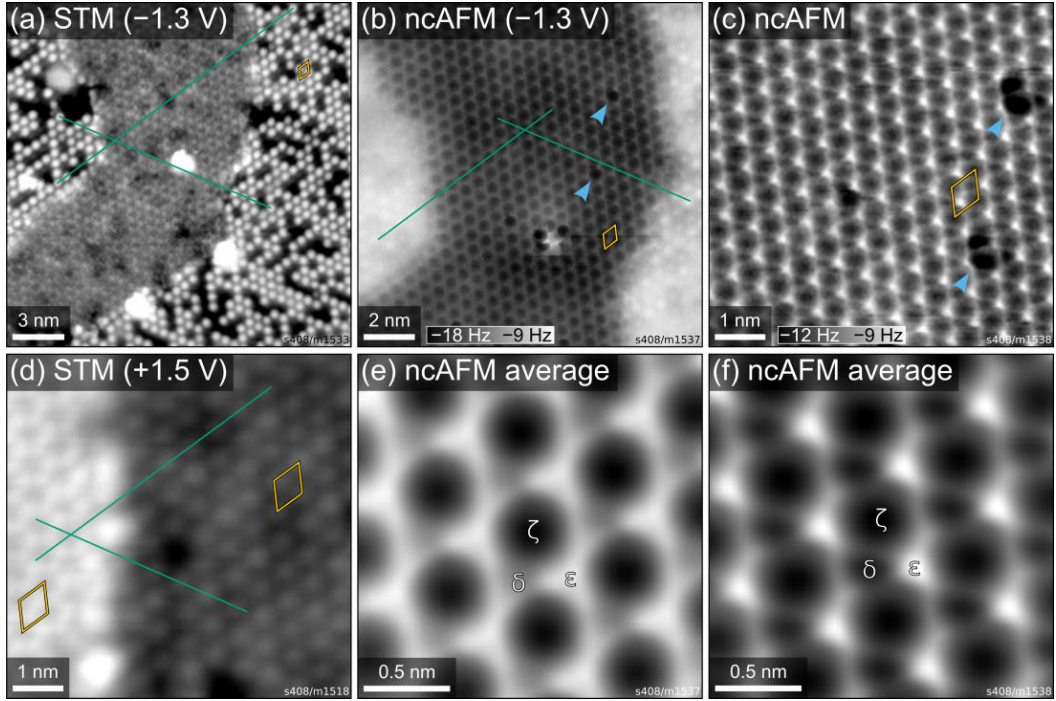

**Figure 5.** Low-temperature ($T = 78$ K) STM and ncAFM images of the "honeycomb" termination (assigned as $Fe_{oct2}$ in previous works) formed under reducing conditions, coexisting with the $Fe_{tet1}$ termination. (a, d) Constant-current STM images acquired at different positions on the sample with $I_{tunnel} = 50$ pA and bias voltages of (a) $U_{sample} = -1.3$ V, (d) $U_{sample} = +1.5$ V. (b, c) Constant-height ncAFM images of the same area as shown in (a) with higher magnification. A sample bias of −1.3 V was applied in (b); the image in (c) was taken without bias at a height 60 pm closer to the surface, with 400 pm oscillation amplitude in both cases. (e, f) Image averages over the unit cells in the honeycomb areas of panels (b) and (c), respectively. Green lines in panels (a), (b) and (d) are aligned with bright features in the $Fe_{tet1}$ areas to highlight relative positions of features in the honeycomb areas. Blue arrows mark the same two defects in (b) and (c). (1 × 1) unit cells are marked in orange. Unit cell corners are placed at $\zeta$ sites, in registry with $Fe_{tet1}$ atoms, both here and in Figure 3.

To obtain slightly reduced surfaces, samples were repeatedly sputtered (1 keV $Ar^+$ ions, 10 min) and annealed in UHV (20 min at 870-930 K), with only the final anneal being performed in $10^{-6}$ mbar $O_2$. After oxygen annealing, the samples were kept at the annealing temperature for another 5 min to ensure low residual oxygen pressure during cooling. This generally resulted in surfaces exposing the $Fe_{tet1}$ termination, as well as patches of another termination with a honeycomb appearance in STM, as shown in Figure 5 (a) and (d). When samples were overly reduced, they also exhibited patches of the reduced "biphase" termination,[14-17] which will not be directly addressed here. The few larger bright features visible in panel (a) are Pt clusters previously used for ncAFM tip preparation,[22] which were subsequently encapsulated during annealing[23] and remained in the subsurface even after more than 10 cycles of sputtering/annealing. The presence of these subsurface clusters does not affect the surface reconstruction outside the clusters' immediate vicinity, as clearly seen in Figure 5 (b). The STM appearance of both the honeycomb and $Fe_{tet1}$ terminations in this dataset are fully consistent with Pt-free data and with other images in the literature.

STM images of the honeycomb termination agree well with previously published results.[11, 13] While the $Fe_{tet1}$ termination is characterized by one bright feature per unit cell, the honeycomb appearance results from two bright features in every unit cell. Point defects consisting of one missing feature are also observed in STM images of the honeycomb phase (Figure S4), as reported previously.[13] Green lines in Figure 5 (a) and (d) highlight the relative positions of bright features in the two terminations. In both cases, the $Fe_{tet1}$ features are aligned with holes in the honeycomb phase. Furthermore, both STM images in Figure 5 show the two terminations at very similar apparent heights, with the $Fe_{tet1}$ phase 25 pm above the honeycomb phase in Figure 5 (a) (filled states, $U_{sample} = -1.3$ V) and 70 pm below the honeycomb phase in Figure 5 (d) (empty states, $U_\text{sample} = +1.5$ V; line profile shown in Figure S4). Both of these values are much smaller than the height expected for a step between terraces of the same termination (485 pm). This bias dependence makes it seem likely that in both cases, the geometric height of the two phases is similar, and that the apparent height difference is caused mainly by differences in the electronic structure. Note however that this does not preclude small differences of the geometric height (*e.g.* additional atoms or a small interlayer distance). These data are in good agreement with the results by Lennie et al., who found the apparent height of the honeycomb phase to be 50 pm above that of the $Fe_\text{tet1}$ phase at +2 V sample bias, and also report alignment of $Fe_\text{tet1}$ features with holes in the honeycomb.[13] Note that this is in conflict with assignment of the honeycomb pattern as an $Fe_\text{oct2}$ termination, which would require the $Fe_\text{tet1}$ features to be aligned with one of the bright features of the honeycomb.

The assignment of a similar height for the two phases is corroborated by the constant-height ncAFM images shown in Figure 5 (b) and (c). Both images were taken on the same sample area as Figure 5 (a). In Figure 5 (b), both the honeycomb and the $Fe_\text{tet1}$ phase are clearly resolved, indicating a similar height. Again, green lines indicate the relative positions of features in the two phases, and $Fe_\text{tet1}$ atoms are in phase with a darker area in the honeycomb. For easier inspection, Figure 5 (e) and (f) show averages over the unit cells of the honeycomb areas in Figure 5 (b) and (c), respectively. This suppresses noise and provides a clear resolution of three different threefold sites of the unit cell in both images, labelled as δ, ε and ζ (following the nomenclature in ref. 13), where ζ is in phase with the $Fe_\text{tet1}$ features. The appearance of ζ and ε as dark and bright is the same in both images, while δ appears with intermediate brightness in Figure 5 (b) but darker in Figure 5 (c), where the tip is closer to the surface.

Overall, both the DFT and microscopy results show that the previous assignment of the honeycomb pattern as an $Fe_{oct2}$ termination is incorrect. We find significantly lower surface energies for alternative terminations (Figure 2), and the alignment of the two phases in STM and ncAFM images (Figure 5) is in conflict with their interpretation as $Fe_{tet1}$ and $Fe_{oct2}$. This registry mismatch is already apparent in STM images published in previous studies.[11, 13] Some uncertainty had previously remained because true atomic heights cannot be accurately determined from the apparent heights in STM, but this ambiguity is removed by ncAFM, which rules out a large step between the $Fe_{tet1}$ and honeycomb terminations in Figure 5. We therefore conclude that any viable model for the reduced termination must have the $Fe_{tet1}$ sites aligned with holes of the honeycomb pattern.

We have introduced two models that fit this structural criterion, presented in Figure 3 (e-h). First, the $Fe_{oct2}$ termination can be modified by introducing one surface oxygen vacancy [Figure 3 (c, d)], then shifting the registry of the surface layer [Figure 3 (e, f)]. We find this shift to be energetically unfavorable, but the energy difference found by DFT (2-5 meV/Å$^2$, Figure S1) is too small to conclusively rule out the possibility. Since the registry shift moves the $Fe_{tet1}$ atom away from its original position and leaves that site empty, the shifted structure is consistent with the observed alignment of $Fe_{tet1}$ features with holes in the honeycomb. In contrast, without the registry shift, the $Fe_{oct2}+V_O$ termination leaves the lateral positions of surface iron atoms with respect to the $Fe_{tet1}$ termination unchanged [Figure 3 (c, d)], and is therefore still in conflict with the scanning-probe images.

The second viable model is a relaxed $Fe_{oct1}+O$ termination, which is favorable at higher oxygen chemical potential [Figure 3 (g, h)]. Here, the surface contains three symmetry-equivalent iron atoms per unit cell, which are brought into a near-perfect tetrahedral coordination with two capping oxygen atoms. Unlike the other models, bright features in the honeycomb pattern seen in STM would here be associated with surface oxygen, rather than iron. Simulated STM images of the $Fe_{oct1}$+O termination (Figure S5) confirm this assignment. The point defects observed in STM (Figure S4) would then most likely correspond to oxygen vacancies. After relaxation, the topmost oxygen atom in the $Fe_{oct1}$+O model is at almost the same height as the surface iron atom in the $Fe_{tet1}$ termination ($\Delta z = 0.17$ Å, see Table S1), in good agreement with the appearance in STM and ncAFM.

While the scanning-probe images in Figure 5 could be rationalized by either the $Fe_{oct1}$+O or the registry-shifted $Fe_{oct2}$+$V_O$ termination, the oxygen-capped $Fe_{oct1}$+O model is more plausible based on other experimental evidence. First, its predicted stability region is adjacent to that of the $Fe_{tet1}$ surface. If the honeycomb termination would correspond to the $Fe_{oct2}$+$V_O$ termination, then it should be possible to also observe separate regions of the $Fe_{oct1}$+O termination, i.e. two different honeycomb patterns. This does not appear to be the case, which suggests that the honeycomb phase corresponds to the model that is stable at a higher oxygen chemical potential. Second, there is previous experimental evidence that the honeycomb regions are much less reactive to adsorbates than the $Fe_{tet1}$ surface.[24, 25] This would agree well with the oxygen-terminated $Fe_{oct1}$+O model, in which all surface iron is fully 4-fold coordinated. In contrast, higher reactivity than the $Fe_{tet1}$ surface would be expected for any $Fe_{oct2}$ or $Fe_{oct2}$+$V_O$ termination, since these expose two under-coordinated iron atoms per unit cell. CO stretching frequencies for adsorption on the $Fe_{oct2}$ ion have been calculated,[8] but were never observed in infrared reflection absorption spectroscopy (IRAS) experiments,[5, 8] suggesting that such sites do not exist or do not accommodate CO. Finally, three distinct features δ, ε and ζ are observed in ncAFM images (Figure 5). This contrast can be interpreted as interaction with two surface atoms at different heights positioned at δ and ε, and no interaction at the ζ site. This fits the two capping oxygen atoms in the $Fe_{oct1}$+O model, which are clearly at different heights. On the other hand, the surface iron atoms of the $Fe_{oct2}$+$V_O$ surfaces are at very similar heights, and one would expect similar contrast in ncAFM. Therefore, we conclude that the honeycomb regions are best explained by the $Fe_{oct1}$+O model.

A further attractive aspect of the improved models for reduced surface terminations is that this shifts the transition point between the $Fe_{tet1}$ surface and the best reduced model to higher oxygen chemical potential. In the surface phase diagram shown in Figure 2, the transition is predicted at $\Delta\mu_O = -2.6$ eV. This is still quite low, but achievable by UHV annealing, unlike the $-3.0$ eV required for a transition to $Fe_{oct2}$. The model therefore helps to understand why patches of the honeycomb phase are commonly observed when flashing or post-annealing samples after oxygen has been pumped out, which puts the sample at a low but somewhat ill-defined chemical potential.

It is important to note that a monophase termination of $Fe_3O_4$(111) with the honeycomb structure cannot be prepared. When samples are reduced further, they instead restructure into the so-called "biphase" termination. However, the new motifs identified here may also be helpful in explaining the constituent structures of the biphase. In our $Fe_{oct2}$+$V_O$ models [Figure 3 (c-f)], the kagome $Fe_{oct1}$ layer is transformed to a tetrahedral coordination, with only one bond per iron atom to the $Fe_2O_3$ layer. This allows for significant flexibility in the placement of the ad-layer, as evidenced by the low energy cost of the registry shift. We tentatively propose that this same tetrahedrally coordinated kagome layer could support either a range of different reduced (1 × 1) structures,[14-17] or an FeO-like adlayer in a moiré structure.[18] Following the 18:17 relationship between substrate and adlayer proposed by Spiridis et al.,[18] a (17 × 17) supercell of a wüstite-based FeO or $Fe_2O_2$ layer could be attached to a (9 × 9) supercell of the $Fe_3O_4$(111) surface with only ≈ 3% lattice strain of the adlayer. The apparently flexible bond angles of tetrahedrally coordinated iron in the kagome $Fe_{oct1}$ layer could conceivably accommodate such an attachment. Of course, significant modifications of such structures and variations in stoichiometry may be necessary to explain the range of different morphologies observed for biphase structures.[14-18] If the "biphase" termination does in fact contain a kagome layer with reduced coordination to oxygen, it may be interesting to investigate whether the weaker linking between iron atoms gives rise to clean kagome bands.[26]

In conclusion, a combination of DFT calculations and scanning-probe methods have allowed us to shed new light on the structural motifs observed on the $Fe_3O_4$(111) surface at reducing conditions. Both the experimental evidence and DFT results conclusively rule out the previously accepted $Fe_{oct2}$ model. Somewhat counterintuitively, we conclude that an oxygen-terminated reconstruction is formed in reducing conditions, which helps explain the relatively inert behavior observed in experiment.

**Experimental and computational methods**

Experiments were performed on natural single crystals (SurfaceNet GmbH, <0.3° miscut). Samples were cleaned by cycles of 1 keV $Ar^+$ or $Ne^+$ sputtering and annealing in oxygen and UHV as described in the main text until free from contaminants as judged by x-ray photoelectron spectroscopy (XPS). Three UHV setups were used in this study: Room-temperature STM was performed in a UHV setup equipped with a nonmonochromatic Al K$\alpha$ X-ray source (VG), a SPECS Phoibos 100 analyzer for XPS, and an Omicron $\mu$-STM. Low-temperature STM and ncAFM were performed in a second setup using an Omicron LTSTM equipped with a Qplus sensor and an in-vacuum preamplifier.[27] Finally, to confirm that the single crystal surfaces are equivalent to thin films studied in previous work, high-quality TPD and XPS data were acquired in a molecular beam setup designed to study the reactivity of oxide single crystals, described in detail in ref. [28]. Samples studied in the ncAFM chamber, which is not equipped with XPS, were first examined in the room-temperature STM chamber, then cleaned again after transfer. Scanning probe images were corrected for distortion and creep of the piezo scanner, as described in ref. [29]. Image averages in Figure 5 were obtained by algorithmically detecting each ζ site of the honeycomb pattern, then averaging over 2×2 nm² image areas centered at those sites.

The Vienna ab initio Simulation Package (VASP)[30, 31] was used for all calculations, with near-core regions described by the projector augmented wave method.[32, 33] A Γ-centered $k$-mesh of $7 \times 7 \times 1$ was used for all (1 × 1) slabs and the plane wave basis set cutoff energy was set to 550 eV. Calculations were performed at the PBE+$U$ level,[34, 35] with an on-site Coulomb repulsion term $U_{eff}$ = 3.61 eV based on previous work.[36] Slabs were relaxed until the residual forces acting on ions were smaller than 0.02 eV/Å. Surface phase diagrams were derived following the approach described by Reuter and Scheffler,[37] using bulk $Fe_3O_4$ and a free oxygen molecule in the triplet state as references. Simulated STM images were created using the Tersoff-Hamann approximation in constant-height mode.[38] The charge states of iron cations were evaluated using the Bader approach.[39-41] Reported Bader charge values are the differences between the 8 valence electrons considered in the calculations and the total projected charge in the Bader volume.

Slabs were constructed from an experimentally determined bulk unit cell (Fd$\overline{3}$m, $a$ = 8.396 Å, JCPDS file[42] 19-629). We primarily used asymmetric slabs containing 15-17 Fe layers depending on the surface termination, with a vacuum gap of at least 15 Å and applying dipole corrections as implemented in VASP. An $Fe_{oct2}$ termination was used at the bottom of the slab, such that the $Fe_{tet1}$-terminated slab is stoichiometric overall. The bottom 7 Fe layers and corresponding oxygen were kept fixed. Slabs yielding the lowest surface energies were also recalculated based on a PBE+$U$-optimized bulk, which overestimates the lattice constant by ≈0.7%. Similarly, we also tested the most relevant terminations on symmetric slabs (13-17 Fe layers). Relative surface energies changed slightly in both cases, but did not significantly affect the conclusions. A comparison of the surface phase diagrams based on the three different setups is shown in Figure S1.

## ASSOCIATED CONTENT

**Supporting Information**. The following files are available free of charge.

Additional figures: surface phase diagrams for different theoretical slab setups, illustration of the relationship between $Fe_{tet1}$ and $Fe_{oct1}+V_O$ terminations, water TPD spectra, STM image showing point defects in the honeycomb termination, and STM simulations. Table listing relative geometric heights of surface atoms in various terminations (PDF).

DFT-optimized structure files (CIF).

## AUTHOR INFORMATION

**Notes**

The authors declare no competing financial interests.

## ACKNOWLEDGMENT

This work was supported by the Austrian Science Fund (FWF) under project number F81, Taming Complexity in Materials Modeling (TACO) (MM, GSP, UD, MS, CF). GSP, FK, JH and MM acknowledge funding from the European Research Council (ERC) under the European Union's Horizon 2020 research and innovation programme (grant agreement No. [864628], Consolidator Research Grant 'E-SAC'). UD, JH and JB acknowledge funding from the European

Research Council (ERC) under the European Union's Horizon 2020 research and innovation programme (grant agreement No. [883395], Advanced Research Grant 'WatFun').

**Supporting Information**

# Oxygen-Terminated (1 × 1) Reconstruction of Reduced Magnetite $Fe_3O_4$(111)

*Florian Kraushofer[1,†], Matthias Meier[1,2], Zdeněk Jakub[1,‡], Johanna Hütner[1], Jan Balajka[1], Jan Hulva[1], Michael Schmid[1], Cesare Franchini[2,3], Ulrike Diebold[1], Gareth S. Parkinson[1]\**

[1] Institute of Applied Physics, Technische Universität Wien, Wiedner Hauptstraße 8-10/E134, 1040 Wien, Austria

[2] University of Vienna, Faculty of Physics and Center for Computational Materials Science, 1090 Wien, Austria

[3] Alma Mater Studiorum, Università di Bologna, 40127 Bologna, Italy

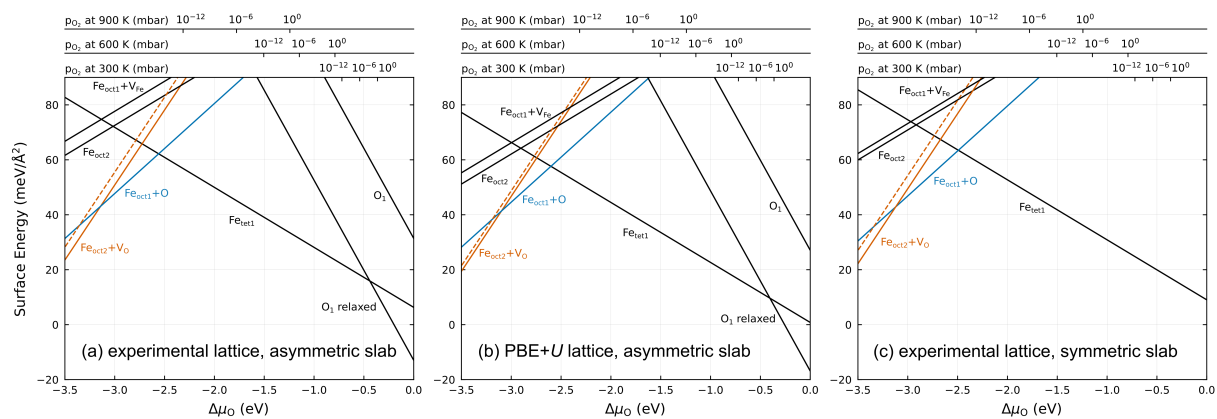

Figure S1: Surface energies of different terminations as a function of the oxygen chemical potential, obtained for different slabs. (a) Asymmetric slab using the experimental lattice constant, also shown as Figure 2 in the main manuscript. (b) Asymmetric slab based on a PBE+$U$-optimized bulk cell. (c) Symmetric slab using the experimental lattice constant. The top axes indicate the pressures corresponding to the chemical potential at the given temperatures. Colored lines are new models introduced in this work, black lines correspond to terminations considered in previous work. "$O_1$ relaxed" is the modified $O_1$ termination introduced in ref. 1, while all other models can be found in ref. 2.

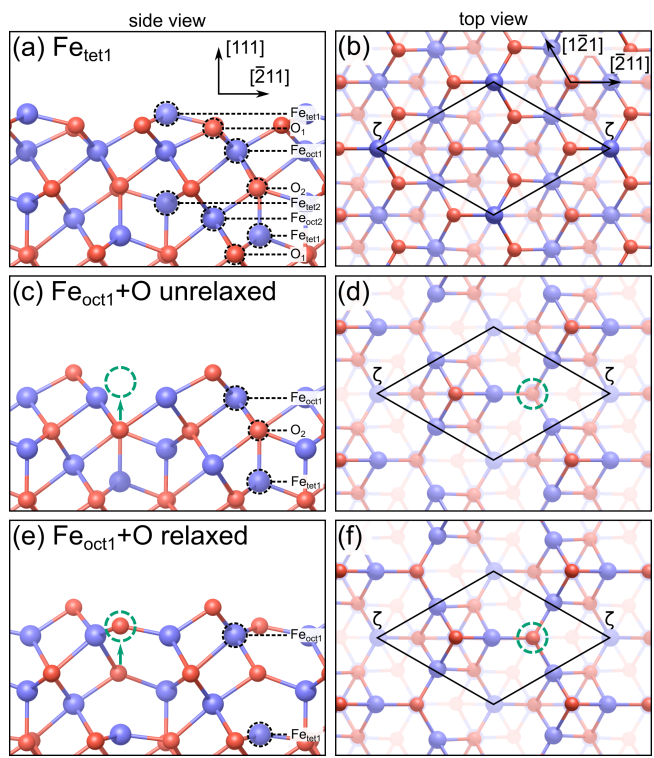

Figure S2: Illustration of the relationship between the Fe$_{tet1}$ and the Fe$_{oct1}$+O terminations. Iron is blue (large), oxygen is red (small). (a, b) The "standard" Fe$_{tet1}$ termination. (c, d) The Fe$_{oct1}$ termination with iron trimers capped by an additional oxygen atom per unit cell, as cut from the bulk structure. (e, f) The same Fe$_{oct1}$+O termination after DFT relaxation, as shown in Figure 3. One subsurface oxygen atom per unit cell breaks its bond to an underlying Fe$_{tet1}$ and relaxes out of the surface, as indicated by the green arrows and dashed green circles in panels (c-f). The unit cell position relative to the bulk is the same in panels (b), (d), and (f), with the corners at Fe$_{tet1}$ sites ($\zeta$ sites).

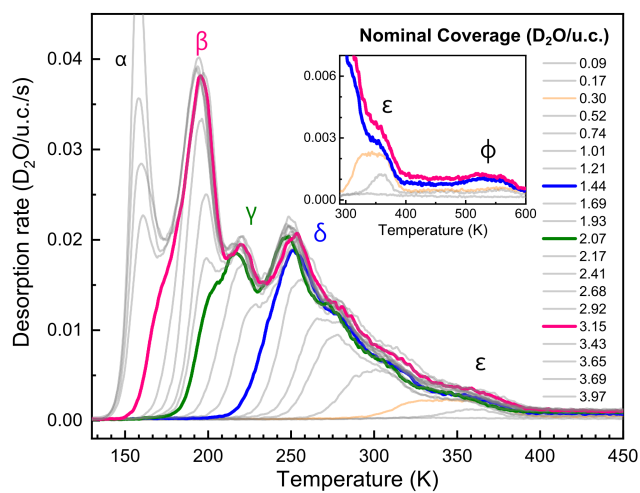

Figure S3: Experimental TPD spectra (1 K/s heating rate) obtained for initial D$_2$O coverages ranging from 0 to 4 molecules per Fe$_{tet1}$-terminated Fe$_3$O$_4$(111)-(1 × 1) unit cell. Water TPD was acquired to confirm that our preparation of single crystal surfaces yields the same Fe$_{tet1}$ termination as the thin film growth reported in ref. 3. A magnified view of desorption peaks $\varepsilon$ and $\Phi$ at higher temperatures is shown in the inset. The colored curves indicate the coverages at which a particular desorption feature saturates. The nominal coverage given in the figure legend may underestimate the actual coverage by up to 10%.

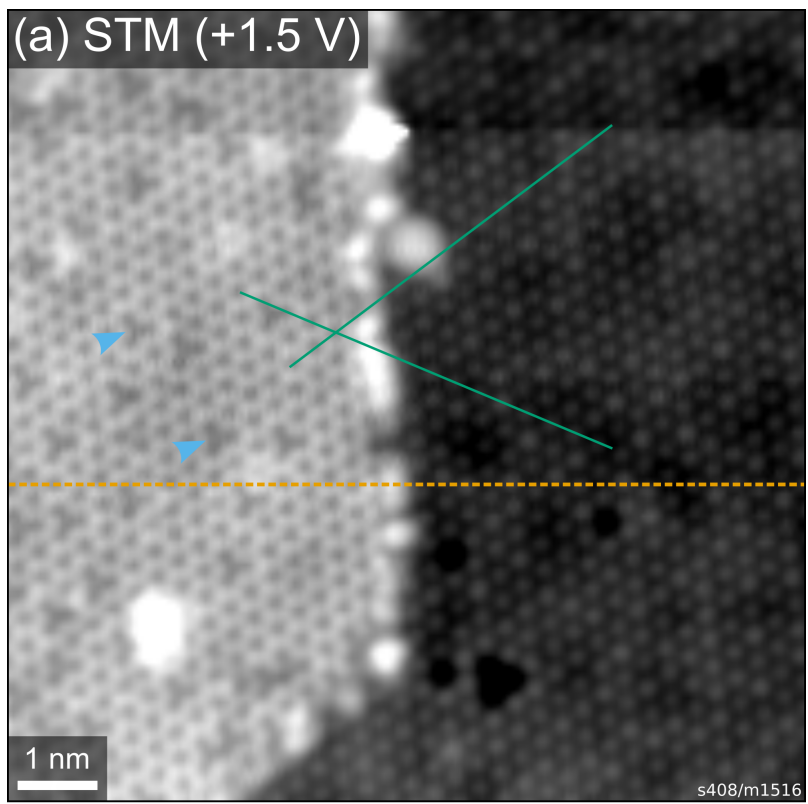

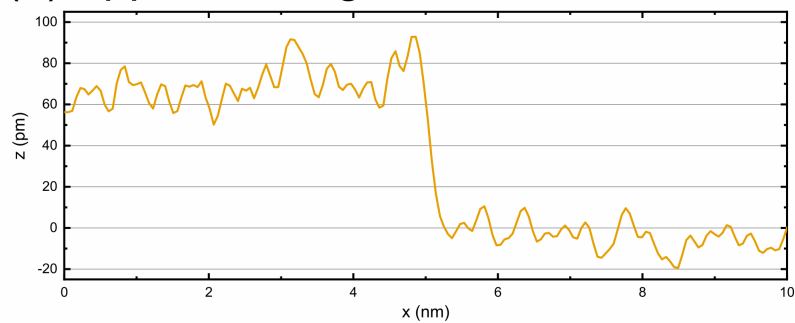

Figure S4: (a) Low-temperature ($T$ = 78 K) STM image ($I_{tunnel}$ = 50 pA, $U_{sample}$ = +1.5 V) of the "honeycomb" termination formed under reducing conditions, coexisting with the Fe$_{tet1}$ termination. Green lines are aligned with bright features in the Fe$_{tet1}$ areas to highlight relative positions of features in the honeycomb areas. Blue arrows mark point defects in the honeycomb phase, consisting of one missing bright feature in the ε site.
(b) Apparent-height profile along the dashed orange line in (a).

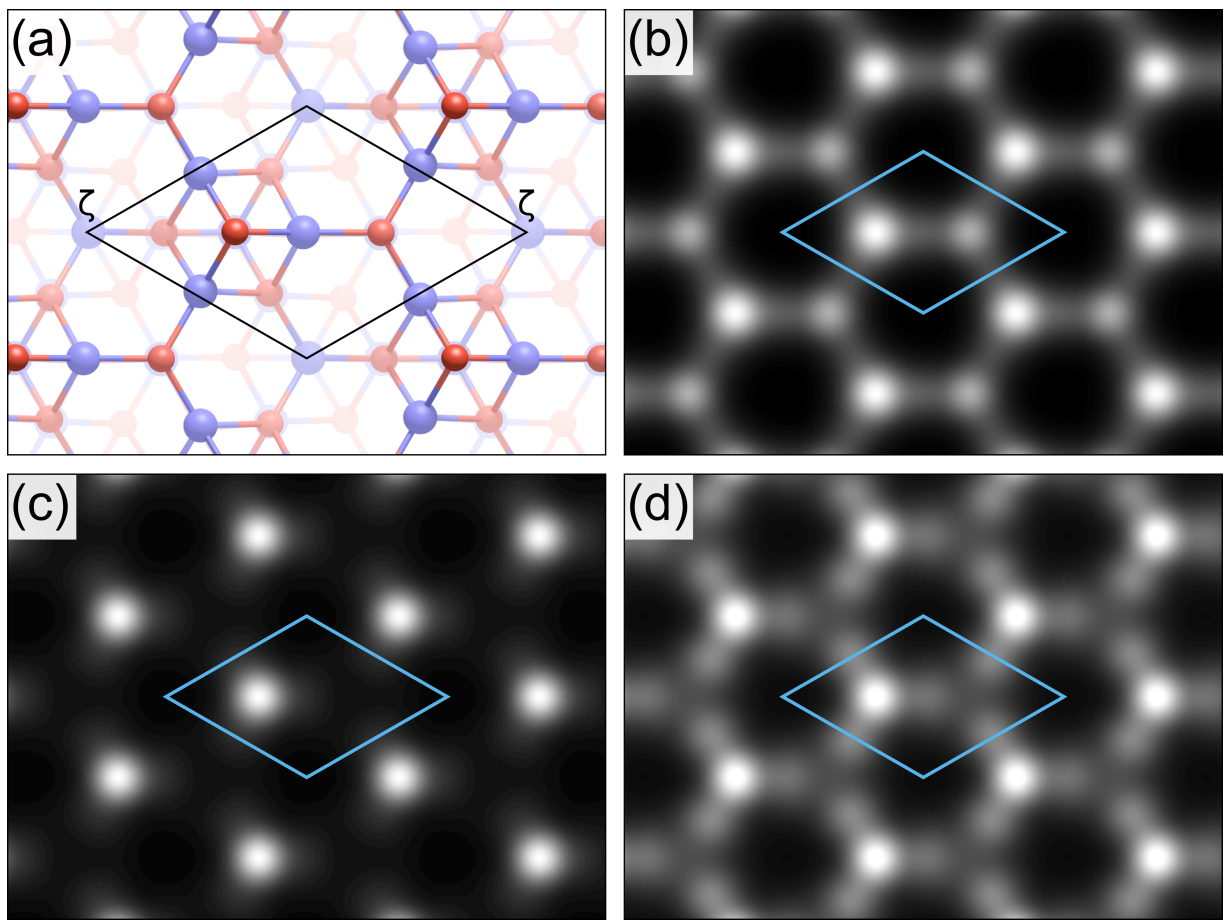

Figure S5: (a) Relaxed Fe$_{oct1}$+O termination as shown in Figure 3 (h) and Figure S2 (f), with the unit cell drawn in black and unit cell corners at Fe$_{tet1}$ sites (site ζ in Figure 5). (b-d) Simulated constant-height STM images of the Fe$_{oct1}$+O termination with sample bias voltages of (b) −2 V, (c) +2 V and (d) +0.5 V. The position and orientation of the blue unit cells is the same as for the black unit cell in (a). The typical honeycomb contrast is reproduced in (b) and (d), while (c) shows only one feature per unit cell (again out of registry with the Fe$_{tet1}$ sites), as is also sometimes observed in experiment.[4,5] Bright features correspond to oxygen atoms in (b) and (c), while both iron and oxygen atoms contribute to the rings in (d).

| Δz [Å] | Fe$_{tet1}$ | Fe$_{oct2}$+V$_O$ | Fe$_{oct2}$+V$_O$ shifted | Fe$_{oct1}$+O |
|---|---|---|---|---|
| surface Fe | 0.00 | +0.78 / +0.40 | +0.83 / +0.78 | −1.24 |
| surface O | −0.43 | +0.27 | +0.41 | −0.17 / −0.87 |

Table S1: Relative height (coordinate perpendicular to the surface) in Å with respect to surface iron atom in the Fe$_{tet1}$ termination for surface iron and oxygen atoms in selected terminations.